\documentclass[useAMS,usenatbib,usegraphicx]{mn2e}

\usepackage[scriptsize]{subfigure}
\usepackage{amssymb,amsmath,bm}
\title[Autoencoder of stellar spectra]{An autoencoder of stellar spectra and its application in automatically estimating atmospheric parameters}
\author[Tan Yang, Xiangru Li]{Tan Yang, Xiangru Li\thanks{E-mail:
xiangru.li@gmail.com (X. Li)}\\
School of Mathematical Sciences, South China Normal University, No. 55, West of Yat-sen Avenue, Guangzhou, 510631, China}
\begin{document}


\pagerange{\pageref{firstpage}--\pageref{lastpage}} \pubyear{2015}

\maketitle

\label{firstpage}

\begin{abstract}
This article investigates the problem of estimating stellar atmospheric parameters from spectra. Feature extraction is a key procedure in estimating stellar parameters automatically. We propose a scheme for spectral feature extraction and atmospheric parameter estimation using the following three procedures: firstly, learn a set of basic structure elements (BSE) from stellar spectra using an autoencoder; secondly, extract representative features from stellar spectra based on the learned BSEs through some procedures of convolution and pooling; thirdly, estimate stellar parameters ($T_\texttt{eff}$, log$~g$, [Fe/H]) using a back-propagation (BP) network. The proposed scheme has been evaluated on both real spectra from Sloan Digital Sky Survey (SDSS)/Sloan Extension for Galactic Understanding and Exploration (SEGUE) and synthetic spectra calculated from Kurucz's new opacity distribution function (NEWODF) models.  The best mean absolute errors (MAEs) are 0.0060 dex for log$~T_\texttt{eff}$, 0.1978 dex for log$~g$ and 0.1770 dex for [Fe/H] for the real spectra and 0.0004 dex for log$~T_\texttt{eff}$, 0.0145 dex for log$~g$ and 0.0070 dex for [Fe/H] for the synthetic spectra.
\end{abstract}


\section{Introduction}\label{Sec:Introduction}
With the development of astronomical observation technology, more and more large-scale sky survey projects have been proposed and implemented: for example, the Sloan Digital Sky Survey \citep[SDSS:][]{York2000,Ahn2012}, the Large Sky Area Multi-Object Fibre Spectroscopic Telescope \citep[LAMOST/Guoshoujing Telescope:][]{Zhao2006,Cui2013,Journal:Luo:2015} and the \emph{Gaia}-ESO Survey  \citep{Journal:Gilmore:2012,Journal:Randich:2013}. Large sky survey projects can automatically observe and collect mass astronomical spectral data. However, the speed of manually processing these massive, high-dimensional data cannot adapt to the rapid growth of data. Therefore, automatic processing and analysis methods are urgently needed for astronomical data \citep{Journal:Zhang:2009}. Because of this, automatic estimation of stellar atmospheric parameters from spectra is currently a hot topic \citep{Recio-Blanco2006, Journal:Fiorentin:2007, Jofre2010, Wu2011, Liu2014}.

Estimating the stellar parameters ($T_\texttt{eff}$, log$~g$, [Fe/H]) from spectra can be abstracted to a process of finding a mapping from a stellar spectrum to its parameters. If there is a set of representative stellar spectra with known parameters (training data), a supervised machine learning method can learn the mapping from training data: for example nearest neighbor (NN), artificial neural networks (ANN) and support vector machines (SVM). Based on the learned mapping, we can estimate the atmospheric parameters for a new spectrum.

An astronomical spectrum usually consists of thousands of fluxes and it is necessary to extract a small number of features to estimate stellar parameters accurately (Section \ref{Sec:Experiments:balance}). Orthogonal transforms such as wavelet transforms (WTs) and principal-component analysis (PCA) are popular choices for extracting features. They give a complete representation of data by a set of complete orthogonal bases. By selecting a small number of transform coefficients as features, dimensionality reduction is reached. The wavelet coefficients of a spectrum are used to estimate stellar parameters \citep{Manteiga2010, Lu2012,Lu2013} and classify spectra \citep{Guo2004,Xing2006}. As the most popular dimension-reduction method, PCA has also been widely used in astronomical data analysis \citep{Whitney1983,Bailer-Jones1998, Zhang2005, Singh2006, Zhang2006, Rosalie2010}. Many parameter-estimation schemes use the projections of data/spectra on some principal components directly as spectral features and have achieved good results \citep{Zhang2005,Zhang2006,Singh2006, Rosalie2010}.

The motivation of our research is to explore the feasibility of predicting atmospheric parameters ($T_\texttt{eff}$, log$~g$, [Fe/H]) from a small number of descriptions computed from information within a local wavelength range of a spectrum. For convenience, this kind of description is referred to as `local and sparse features'. This characteristic of local representation helps to trace back the potential spectral lines effective in estimation (See section \ref{Sec:Feature Extract:Discussion}).

This work proposes a scheme based on autoencoders, convolution and pooling techniques to extract local and sparse features. Autoencoder and convolution operations give a statistical non-orthogonal decomposition, which leads to a redundant and overcomplete representation of data. The `overcomplete' representation (equation \ref{Equ:convolution}) of a spectrum has many more dimensions than the original spectrum (equation \ref{Equ:AE:spectrum}). While complete representations based on orthogonal transforms are mature and popular feature extraction methods, the redundant, `overcomplete' representations \citep{Olshausen2001, Teh2003} have been advocated and used successfully in many fields. In literature, researches \citep{Journal:Yee:2003} show that redundancy and overcompleteness help in computing some features in subsequent procedures, with improved robustness in the presence of noise \citep{Journal:Simoncelli:1992}, more compactness and more interpretability \citep{Jouranl:Mallat:1993}.

From the overcomplete representation, a sparse representation (equation \ref{Equ:pooling}) is computed using pooling and maximization operations. The pooling and maximization operations are actually a competition strategy. In this procedure, much redundancy is removed by the competitions between multiple redundant components. This competition helps to restrain the copies with more noises and a robust representation is obtained. A typical advantage of this scheme is that it can express many suitable and meaningful structures in data in some applications. It is shown that this scheme does extract some meaningful local features (section \ref{Sec:Feature Extract:Discussion}) for automatically estimating atmospheric parameters (Section \ref{Sec:Experiments}).

The rest of this article is organized as follows: Section \ref{Sec:Data} describes the spectra used in this work.  Section \ref{Sec:Framework} presents the framework of the proposed scheme. Section \ref{Sec:LearningBSE} introduces the autoencoder network, the concept of basic structure elements (BSEs) and the BSE learning method using autoencoders. Sections \ref{Sec:Feature Extract} and \ref{Sec:Estimator} describe the proposed feature-extraction method based on BSEs and the estimating method back-propogation (BP) network, respectively. Section \ref{Sec:Configuration} investigates the optimization of the proposed scheme. Section \ref{Sec:Experiments} reports some experimental evaluations and discusses the rationality and robustness of the proposed scheme. Section \ref{Sec:Conclusion} concludes this work.

\section{DATA SETS AND PREPROCESSING}\label{Sec:Data}

\subsection{Real spectra}\label{Sec:Data:SDSS}
In this article, we use two spectral sets to evaluate the proposed scheme. 5000 stellar spectra selected from SDSS-DR7 \citep{Journal:Abazajian:2009} compose the real spectrum set. Each spectrum has 3000 fluxes in a logarithmic wavelength range [3.6000, 3.8999] with a sampling resolution of 0.0001. The ranges of the three parameters are [4163, 9685]~K for $T_\texttt{eff}$,  [1.260, 4.994]~dex for log$~g$ and [-3.437, 0.1820]~dex for [Fe/H].

\subsection{Synthetic spectra}\label{Sec:Data:synthetic}
A set of 18~969 synthetic spectra is generated with Kurucz's new opacity distribution function (NEWODF) models \citep{Castelli2003} using the package SPECTRUM \citep{Gray1994} and 830~828 atomic and molecular lines. The solar atomic abundances we used are derived from \citet{Grevesse1998}. The grids of the synthetic stellar spectra span the parameter ranges [4000, 9750] K in $T_\texttt{eff}$ (45 values, step size 100~K between 4000 and 75~00 K and 250~K between 7750 and 9750~K), [1, 5] dex in log$~g$ (17 values, step size 0.25 dex) and [-3.6, 0.3] dex in [Fe/H] (27 values, step size 0.2 between -3.6 and -1 dex, 0.1 between -1 and 0.3 dex).

\subsection{Preprocessing and data partitioning}\label{Sec:sub:preprocessing}

 The proposed scheme is implemented based on ANN. Usually an ANN requires that each input component has been normalized to eliminate impacts on input data resulting from range differences from flux to flux. Suppose that
\begin{equation}\label{Equ:AE:spectrum}
\bm{x}=(x_1,\cdots,x_l)^T
\end{equation}
is a spectrum. The normalized spectrum $\bar{\bm{x}}$ is calculated by
\begin{equation}\label{Equ:AE:normalized}
\bar{\bm{x}}=\frac{\bm{x}}{\|\bm{x}\|_2},
\end{equation}
where $\|\bm{x}\|_2 = (\sum_{i=1}^{l}{x_i^2})^{0.5}$. In addition, to reduce the dynamical range and in order better to represent the uncertainties of spectral data, $T_\texttt{eff}$ is replaced by log$~T_\texttt{eff}$ in both sets \citep{Journal:Fiorentin:2007,Journal:Li:2014}.

The proposed scheme of this work belongs to the class of statistical learning methods. The fundamental idea is to discover the predictive relationships between stellar spectra and atmospheric parameters $T_\texttt{eff}$, log$~g$ and [Fe/H] from empirical data, which constitutes a training set. At the same time, the performance of the predictive relationships discovered should also be evaluated objectively. Therefore, a separate, independent set of stellar spectra is needed for this evaluation, usually called a test set in machine learning. However, most learning methods tend to overfit the empirical data. In other words, statistical learning methods can unravel some alleged relationships from the training data that do not hold in general. In order to avoid overfitting, we need a third independent spectrum set for optimizing the parameters (Section \ref{Sec:Configuration}) of the framework that need to be adjusted objectively when investigating the potential relationships. This third spectrum set and the reference parameters constitutes the validation set.

Therefore, in each experiment, we split the total spectrum samples into three subsets: a training set (60 percent), a validation set (20 percent) and a test set (20 percent). The training set is the carrier of knowledge and the proposed scheme should learn from this set. The validation set is a mentor/instructor of the proposed scheme that can independently and objectively give some advice to the learning process. The training set and the validation set are used in establishing a spectral parameterization, while the test set acts as a referee to evaluate the performance of the established spectral parameterization objectively. The roles of these three subsets are listed in Table \ref{Tab:Data sets}.

\begin{table}
\centering
\caption{ Roles of the three data sets}
\begin{tabular}{lp{0.65\columnwidth}}
\hline
Data set        &  Roles\\
\hline
Training set    & (1) Generate spectral patches to construct a training set for an autoencoder (Subsection \ref{Sec:LearningBSE:Learning})\\
                & (2) Learn in optimizing the configuration (Section \ref{Sec:Configuration})\\
                & (3) Train the BP network (Section \ref{Sec:Estimator})\\
                \hline
Validation set  & Evaluate the performance of the learned spectral parameterization in optimizing the configuration (Section \ref{Sec:Configuration})\\
                 \hline
Test set        &  Evaluate the performance of the proposed scheme (Section \ref{Sec:Experiments}) \\
\hline
\end{tabular}\label{Tab:Data sets}
\end{table}

\section{Framework}\label{Sec:Framework}
\begin{figure}
\centering
\includegraphics[width =1.6in]{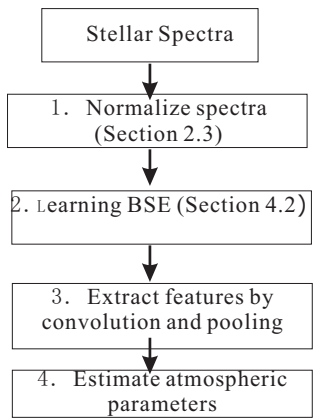}
\caption{A flowchart to show the order of procedures in the proposed scheme. The procedure `Learning BSE' is only used in investigation/training. In application/test stage, we can extract features by convolution and pooling after normalizing spectra using the learned BSE in training.}
\label{Fig:flowchart}
\end{figure}

There are multiple procedures in our researches; a flowchart shown in Fig. \ref{Fig:flowchart} illustrates the end-to-end flow.

Overall, our work can be divided into two stages: (1) a research stage and (2) an application stage. These two stages can be implemented automatically based on the flowchart in Fig. \ref{Fig:flowchart}. As shown in this flowchart, in research stage, we can obtain an optimized configuration for the proposed scheme and a spectral parameterization by which we can map a spectrum approximately to its atmospheric parameters. In the application stage, we can compute the atmospheric parameters from a spectrum. More about optimization of the configuration is discussed in Section \ref{Sec:Configuration}.

This work used multiple acronyms and notations. To facilitate readability, we summarize them in Table \ref{Tab:acronyms_notations}.

\begin{table*}
\centering
\caption{Acronyms and notations used in this work. AN: acronym or notation.}
\begin{tabular}{p{1.2cm} p{4cm} c c p{4cm} c}
\hline \hline
AN         &  ~~~~~~Meaning             &   First appearance   &
AN         &  ~~~~~~Meaning             &   First appearance    \\ \hline
$a^{(l)}_j(\bm{x})$ & the output of the $j$th node in the $l$th layer for an input $\bm{x}$ & Section \ref{Sec:LearningBSE:Autoender} &
$ap$                & an atmospheric parameter log$~T_\texttt{eff}$, log$~g$ or [Fe/H]& Section \ref{Sec:Configuration} \\
$b^{(l)}_j$         & the bias of the $j$th node in the $(l+1)$th layer & Section \ref{Sec:LearningBSE:Autoender}       &
$\beta$             & a regularized parameter in the objective function $J(\cdot,\cdot)$  &Section \ref{Sec:LearningBSE:Autoender} \\
BSE                 & basic structure element                                   & Abstract                       &
DCP                 & description by convolution and pooling                     & Section \ref{Sec:Feature Extract:Pooling}\\
$f(\cdot)$          & an estimation method & Section \ref{Sec:Estimator}                                           &
$g(\cdot)$          & an activation function of an autoencoder & Section \ref{Sec:LearningBSE:Autoender}          \\
$h_{W,b}$           & the mapping of an autoencoder  & Section \ref{Sec:LearningBSE:Autoender}              &
$J(\cdot,\cdot)$    & the objective function of an autoencoder & Section \ref{Sec:LearningBSE:Autoender}            \\
$\mathbf{KL}(\cdot\parallel\cdot)$  & relative entropy of the average output of a hidden node and its expected output  & Section \ref
{Sec:LearningBSE:Autoender}                         &
$\lambda$           & a weight decay parameter in the objective function $J(\cdot, \cdot)$ & Section \ref{Sec:LearningBSE:Autoender} \\
$\hat{\lambda}$     & optimized value for $\lambda$. $\hat{\beta}$, $\hat{\rho}$, $\hat{n}$, $\hat{N}_p$, $\hat{n}^{BP}_{hl}$ and $\hat{\bm{n}}^{BP}_{nhl}$ are defined similarly                 & Section \ref{Sec:Configuration}                      &
$m$                 &  number of nodes in the input layer of an autoencoder & Section \ref{Sec:LearningBSE:Autoender}  \\
$n$                 & number of nodes in the hidden layer of an autoencoder  & Section \ref{Sec:LearningBSE:Autoender} &
$n^{BP}_{hl}$       & number of hidden layers in a BP network                  & Section \ref{Sec:Estimator}        \\
$\tilde{n}^{BP}_{hl}$  & initialized value of $n^{BP}_{hl}$                  & Section \ref{Sec:Configuration}    &
$\bm{n}^{BP}_{nhl}$ & number of nodes in the hidden layers of a BP network     & Section \ref{Sec:Estimator}        \\
$\tilde{\bm{n}}^{BP}_{nhl}$ & an initialization of $\bm{n}^{BP}_{nhl}$                & Section \ref{Sec:Configuration} &
$N$                 & number of samples in a data set $S$ or $S^{tr}$ depending on its context & Section \ref{Sec:LearningBSE:Autoender}            \\
$N_p$               & number of pools                                         & Section \ref{Sec:Feature Extract:Pooling} &
$\rho$              & an desired activation level                             &Section \ref{Sec:LearningBSE:Autoender}    \\
$\bar{\rho}_j$      & the average activation of the $j$th hidden node of an autoencoder &Section \ref{Sec:LearningBSE:Autoender}&
$RR_{\lambda}$      & restricted (search) range for $\lambda$. $RR_{\beta}$, $RR_{\rho}$, $RR_{n}$, $RR_{{N}_p}$, $RR_{{n}^{BP}_{hl}}$ and $RR_
{{\bm{n}}^{BP}_{nhl}}$ are defined similarly  & Section \ref{Sec:Configuration}                                      \\
$S$                 & a data set with spectra and atmospheric parameters            & Section \ref{Sec:Estimator}       &
$S_{BSE}$           & a set of BSEs                                               & Section \ref{Sec:LearningBSE:Learning}\\
$S^{tr}$            & a training set of stellar spectra                                   & Section \ref{Sec:LearningBSE:Learning} &       $S^{tr\_ae}$        & a training set for an autoencoder, consists of some spectral patches   & Section \ref{Sec:LearningBSE:Autoender}\\
$v^{(j)}_q$         & the maximum convolution response with the $j$th BSE in the $q$th pool & Section \ref{Sec:Feature Extract:Pooling}  &       $\bm{w}$            & description by convolution and pooling (DCP)            & Section \ref{Sec:Feature Extract:Pooling}           \\
$W^{(1)}_{j\cdot}$  & the $j$th BSE        & Section \ref{Sec:LearningBSE:Autoender}                                  &
$W^{(l)}_{ji}$      &a weight between the $i$th node in the $l$th layer and the $j$th node in the $(l + 1)$th layer& Section \ref
{Sec:LearningBSE:Autoender}                                                                                           \\
$\bm{x}$            &  a spectrum or a spectral patch for an autoencoder depending on its context          &  Section \ref{Sec:sub:preprocessing}                                     &
$\bar{\bm{x}}$      & a normalized spectrum &  Section \ref{Sec:sub:preprocessing}                                     \\
$\bm{y}$            &  output of an autoencoder & Section \ref{Sec:LearningBSE:Autoender}                               &
$\bm{z}$                 & a convolution response of a spectrum with BSEs     &Section \ref{Sec:Feature Extract:Convolution}    \\
\hline\hline
\end{tabular}\label{Tab:acronyms_notations}
\end{table*}

\section{Learning BSEs using an autoencoder}\label{Sec:LearningBSE}

\subsection{An Autoencoder}\label{Sec:LearningBSE:Autoender}
An autoencoder is a special kind of ANN, initially proposed as a data dimensionality reduction scheme \citep{Hinton2006} and now is widely used in image analysis \citep{Tan2010,Shin2011} and speech processing \citep{Vishnubhotla2010,Deng2013}.

\begin{figure}
\centering
\includegraphics[width=85mm]{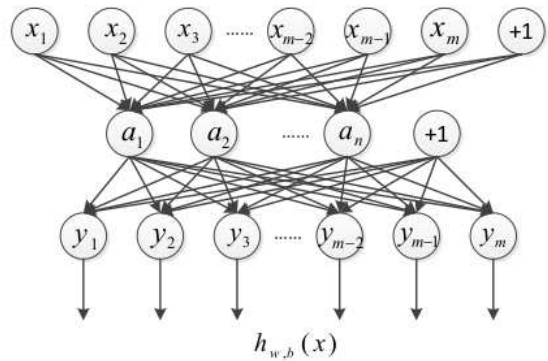}
\caption{A framework of an autoencoder. The number of output nodes is equal to that of input nodes. The learning objective of the network is to make the output $h_{W,b}(x)$ as close as possible to input $\bm{x}$.}
\label{Fig:autoencoder}
\end{figure}

An autoencoder adopts the framework shown in Fig. \ref{Fig:autoencoder} and is usually used to extract features by unsupervised learning.
In an autoencoder, there is only one hidden layer and the number of nodes in the output layer is equal to that in the input layer. The output $y_i$ is an approximation of the corresponding input $x_i$,
\begin{equation}\label{Equ:AE:input_output}
 y_i \approx x_i, i = 1, \cdots, m.
\end{equation}
The learning objective of an autoencoder is to find a set of weights, labeled with lines in Fig. \ref{Fig:autoencoder}, between the nodes in different layers based on a set of empirical data (a training set). In other words, the autoencoder tries to approximate an identity function on the training set.

By setting the number of hidden nodes to be far smaller than that of input nodes, an autoencoder can achieve dimension reduction. For example, when an autoencoder has 200 input nodes and 20 hidden nodes (equivalently, $m = 200$ and $n = 20$ in Fig. \ref{Fig:autoencoder}), the original 200-dimensional input could be `reconstructed' approximately from the 20-dimensional output of the hidden layer. If we use the output of the hidden layer as a representation of an input of the network, the autoencoder plays the role of a feature extractor.

To introduce the implementation details of the autoencoder, we utilize the notations in an online tutorial ``UFLDL Tutorial" \citep{Tutorial:Andrew:2010}. 

In the network of Fig. \ref{Fig:autoencoder}, let the layer labels be 1 for the input nodes, 2 for the hidden nodes and 3 for the output nodes and suppose $W^{(l)}_{ji}$ ($l = 1, 2$) represents a weight between the $i$th node in the $l$th layer and the $j$th node in the $(l+1)$th layer, $b^{(l)}_j$  ($l = 1, 2$) is the bias of the $j$th node in the $(l+1)$th layer and $a^{(l)}_j(\bm{x})$  ($l = 1, 2, 3$) is the output of the $j$th node in the $l$th layer for input $\bm{x} = (x_1, \cdots, x_m)^T$. Then, $a^{(1)}_i(\bm{x}) = x_i$ for the input nodes, $a^{(3)}_k(\bm{x}) = y_k$ for the output nodes.

For compact expressions in equations (\ref{Equ:AE:nodes}) and (\ref{Equ:Object}), we introduce three variables $s_1$, $s_2$ and $s_3$ respectively representing the number of nodes in three layers of an autoencoder network (Fig. \ref{Fig:autoencoder}). Then, $s_1 = m$, $s_2 = n$ and $s_3 = m$ and the relationship between the nodes in different layers is
\begin{equation}\label{Equ:AE:nodes}
a^{(l+1)}_j=g(\sum^{s_l}_{i=1}W^{(l)}_{ji}a^{(l)}_i+b^{(l)}_j), l =1, 2.
\end{equation}
In equation (\ref{Equ:AE:nodes}), the $g(\cdot)$ is referred to in the literature as an activation function. Two common choices for the activation function are a sigmoid function
\begin{equation}\label{Equ:AE:sigmoid}
   g(z) = \frac{1}{1+e^{-z}}
\end{equation}
and a hyperbolic tangent function
\begin{equation}\label{Equ:AE:tangent}
   g(z) = \frac{e^z - e^{-z}}{e^z + e^{-z}}.
\end{equation}
This work uses the sigmoid function in equation (\ref{Equ:AE:sigmoid}).

Overall, the network in Fig. \ref{Fig:autoencoder} implements a non-linear mapping $h_{W,b}(\cdot)$ from an input $\bm{x} = (x_1,\cdots,x_m)^T$ to an output $\bm{y} = (y_1,\cdots,y_m)^T$:
\begin{equation}
    \bm{y} = h_{W,b}(\bm{x}),
\end{equation}
where $W = \{W^{(l)}_{ji}\}$ represents the set of the weights of an autoencoder network and $b = \{b^{(l)}_j\}$ the set of biases.

Suppose that $S^{tr\_ae}$ is a training set for an autoencoder. To obtain the parameters $W$ and $b$ for an autoencoder network based on equation (\ref{Equ:AE:input_output}), we can minimize an objective function, $J$,  in equation (\ref{Equ:Object})
\begin{eqnarray}\label{Equ:Object}
\lefteqn{J(W,b)=\frac{1}{N}\sum_{\bm{x} \in S^{tr\_ae}}(\frac{1}{2}\|h_{w,b}(\bm{x})-\bm{x}\|^2)}\nonumber\\
&&+\frac{\lambda}{2}\sum^2_{l=1}\sum^{s_l}_{i=1}\sum^{s_{l+1}}_{j=1}(W^{(l)}_{ji})^2+\beta\sum^{s_2}_{j=1}\mathbf{KL}(\rho\parallel\bar{\rho}_j
),
\end{eqnarray}
where $\bar{\rho}_j$ is the average output of the $j$th hidden node,
\begin{equation}
\bar{\rho}_j=\frac{1}{N}\sum_{\bm{x} \in S^{tr\_ae}}[a^{(2)}_j(\bm{x})],
\end{equation}
\begin{equation}
  \mathbf{KL}(\rho\parallel\bar{\rho}_j) = \rho \log{\frac{\rho}{\bar{\rho}}} + (1 - \rho) \log\frac{1- \rho}{1- \bar{\rho}_j},
\end{equation}
$N$ is the number of samples in a training set\footnote{The $tr\_ae$ is the abbreviation of a training set for an autoencoder.} $S^{tr\_ae}$ and $\lambda \geq 0$, $\beta\geq 0$ and $\rho > 0$ are three preset parameters of the spectral parameterization. These three preset parameters control the relative importance of the three terms in equation (\ref{Equ:Object}). In the literature, the $\lambda$ is usually referred to as a weight decay parameter.

In equation (\ref{Equ:Object}), the first term represents an empirical error evaluation between the actual output and the expected output of an autoencoder and ensures a good reconstruction performance of the network. The second term, a regularization term of $W^{(l)}_{ji}$, is used to overcome possible overfitting to the training set by reducing the scheme's complexity.

The third term with weighted coefficient $\beta$ is a penalty term for sparsity in outputs of the hidden layer. $\mathbf{KL}(\rho\parallel\bar{\rho}_j)$ is the relative entropy of the average output, $\bar{\rho}_j$, and a desired activation level $\rho$. $\mathbf{KL}(\rho\parallel\bar{\rho}_j)$ increases monotonically with increasing distance between $\bar{\rho}_j$ and $\rho$ and encourages the average activation, $\bar{\rho}$, of the hidden layer to be close to a desired average activation $\rho$.

\subsection{Learning BSEs}\label{Sec:LearningBSE:Learning}
BSEs consist of a set of templates of spectral patches with a limited wavelength range. Using the BSE, we can extract local and sparse features (Section \ref{Sec:Feature Extract}). This work studies the feasibility of learning BSEs through an autoencoder for automatic estimation of atmospheric parameters. In an autoencoder, the weighted sum $\sum^{m}_{i=1}W^{(1)}_{ji}a^{(1)}_i(\bm{x})+b^{(1)}_j$ in the $j$th hidden node is essentially a projection of an input $\bm{x}$ on the vector $W^{(1)}_{j\cdot}=(W^{(1)}_{j1},W^{(1)}_{j2},\cdots,W^{(1)}_{jm})^T$ of weights, which is similar to the coefficients of a vector in a coordinate system. Thus, $W^{(1)}_{j\cdot}$ can be regarded as a ``basis" for the input data and in this article we name them a basic structure element (BSE), where $j = 1,\cdots, n$.

To learn BSEs through an autoencoder, a training set $S^{tr\_ae}$ was constructed to represent the local information of a stellar spectrum. Let $S^{tr}$ be a training set of stellar spectra (Section \ref{Sec:Data}). The BSE training set $S^{tr\_ae}$  is constructed from a spectral training set $S^{tr}$ in the following way:
\begin{enumerate}\setlength{\itemsep}{0pt} \setlength{\parsep}{0pt} \setlength{\parskip}{0pt}
                    \item[(1)] randomly select a spectrum, $\bm{x} = (x_1, \cdots, x_l)$, from $S^{tr}$, where $l>0$ represents the number of fluxes of a spectrum;
                    \item[(2)] randomly generate an integer $j$ satisfying $1 \leq j \leq n-m+1$, where $m>0$ represents the dimension of a sample in $S^{tr\_ae}$ and is consistent with the number of input nodes of the autoencoder in Fig. \ref{Fig:autoencoder};
                    \item[(3)] take $(x_j, x_{j+1}, \cdots, x_{j+m-1})^T$ as a sample of $S^{tr\_ae}$.
\end{enumerate}

By repeating the above three procedures, we obtain a BSE training set  $S^{tr\_ae}$. Therefore, $S^{tr\_ae}$ actually consisits of a series of spectral patches. Considering the widths of lines of a stellar spectrum, $m$ is empirically taken as 81 in this work. In the proposed scheme, 100~000 such patches are generated to constitute the BSE training set. These patches are not generated from a specific wavelength position, therefore $S^{tr\_ae}$ expresses the general structures of all spectral patches with length $m$.

To learn a set of BSEs, we input the generated training set $S^{tr\_ae}$ into an autoencoder (Section \ref{Sec:LearningBSE:Autoender}) and compute a set of spectral templates, BSEs:
\begin{equation}\label{Equ:BSEs}
S_\texttt{BSE} = \{W^{(1)}_{1\cdot}, \cdots, W^{(1)}_{n\cdot} \},
\end{equation}
where every BSE, $W^{(1)}_{j\cdot}$, is a vector representing a basic pattern of spectral patches,
\begin{equation}\label{Equ:BSE}
W^{(1)}_{j\cdot}=(W^{(1)}_{j1},W^{(1)}_{j2},\cdots,W^{(1)}_{jm})^T.
\end{equation}

\section{Feature extraction}\label{Sec:Feature Extract}

This work proposes to extract features by performing convolution and pooling operations on the computed BSEs and a stellar spectrum.

\subsection{Convolution}\label{Sec:Feature Extract:Convolution}
Let
\begin{equation}\label{Equ:spectrum}
  \bm{x}= (x_1,\cdots,x_l)^T
\end{equation}
denote a spectrum. Using the extracted BSEs $S_\texttt{BSE}$ in equation (\ref{Equ:BSEs}) and a convolution operation, we filter the spectrum $\bm{x}$:
\begin{equation}\label{Equ:convolution:ele}
  z_i^{(j)} = \sum_{p = 1}^{m}{ W^{(1)}_{jp} x_{p+i-1} }
\end{equation}
and transform the spectrum $\bm{x}$ into
\begin{equation}\label{Equ:convolution}
   \begin{split}
   \bm{z} =& ({(z^{(1)})}^T,  \cdots, {(z^{(n)})}^T )^T\\
          =& (z_1^{(1)}, \cdots,  z_{l-m+1}^{(1)}, \cdots, z_1^{(n)}, \cdots,  z_{l-m+1}^{(n)} )^T,\\
   \end{split}
\end{equation}
where $i = 1, \cdots, l-m+1, j =1, \cdots, n$. Then,
\begin{equation}
   z^{(j)} = {(z_1^{(j)}, \cdots,  z_{l-m+1}^{(j)})}^T
\end{equation}
is the convolution response vector of the $j$th BSE structure $W^{(1)}_{j\cdot}$ in equation (\ref{Equ:BSEs}).

In this work, a BSE structure template has $m=81$ components (equation \ref{Equ:BSE}) and a SDSS spectrum is represented with $l=3000$ fluxes. Therefore, there are $l-m+1=2920$ convolution responses for any one BSE structure template and spectrum. From the $n  = 25$ BSE structure templates in equation (\ref{Equ:BSEs}), we obtain $(l-m+1)\times n = 2920\times 25$ convolution responses (in equation \ref{Equ:convolution}) for every spectrum.

\subsection{Pooling}\label{Sec:Feature Extract:Pooling}
However, there are many redundancies in the convolution response description, $\bm{z}$ in equation (\ref{Equ:convolution}), of a spectrum for physical atmospheric parameter estimation. Redundancies usually result in overfitting. To overcome this problem, a `pooling' method is adopted to reduce dimension by merging the features of different positions. In this pooling method, we first equally partition the convolution response description into $N_p$ pools according to their wavelength positions, then choose the maximum response in each pool as the final spectral feature:
\begin{equation}\label{Equ:pooling:vec_sub}
  v_q^{(j)} = \max{\{ z_i^{(j)},  \frac{l}{N_p} \times (q - 1) + 1 \leq i \leq \frac{l}{N_p} \times q\}},
\end{equation}
where $q = 1, \cdots, N_p, j =1, \cdots, n$, $\frac{l}{N_p}$ is the length of every pool, $l$ is the number of fluxes in a spectrum and $N_{p}$ is the number of pools. The $v_q^{(j)}$ is referred to as a maximum convolution response. Thus, the spectrum $\bm{x}$ is transformed to
\begin{equation}\label{Equ:pooling}
   \bm{w} = (v_1^{(1)}, \cdots,  v_{N_p}^{(1)}, \cdots, v_1^{(n)}, \cdots,  v_{N_p}^{(n)} ).
\end{equation}
 For convenience, the detected features in equation (\ref{Equ:pooling}) are named description by convolution and pooling (DCP). For a SDSS spectrum, the dimension of its description decreases from $(l-m+1)\times n = 2920\times 25$ (convolution responses) to $N_p \times n = 10 \times 25$ (DCP), if the pool number $N_p$ is 10.

\subsection{Discussions on convolution and pooling}\label{Sec:Feature Extract:Discussion}
Pooling is a form of non-linear downsampling operation, which combines the responses of feature detectors at nearby locations into some statistical variables. The pooling method was proposed in Hubel and Wiesel's work on complex cells in the visual cortex \citep{Hubel1962} and is now used in a large number of applications: for example computer vision and speech recognition. Theoretical analysis \citep{Boureau2010} suggests that the pooling method works well when the features are sparse. In image- and speech-processing fields, convolution and pooling based on local bases has been proven to be an effective feature-extraction method \citep{Nagi2011, Shin2011,Ashish2013}.

In theory, the convolution operation can reduce negative effects from noise and the pooling operation can restrain the negative influence from some imperfections (e.g. sky lines and cosmic ray removal residuals, residual calibration defects) by competing between multiple convolution responses of the extracted BSEs and a spectrum (the maximization procedure). Some effects of convolution and pooling are investigated in the following part of this subsection and further discussed in Sections \ref{Sec:Experiments:balance} and \ref{Sec:Experiments:stiching} based on experimental results.

\begin{figure*}
\centering
\includegraphics[width =5in]{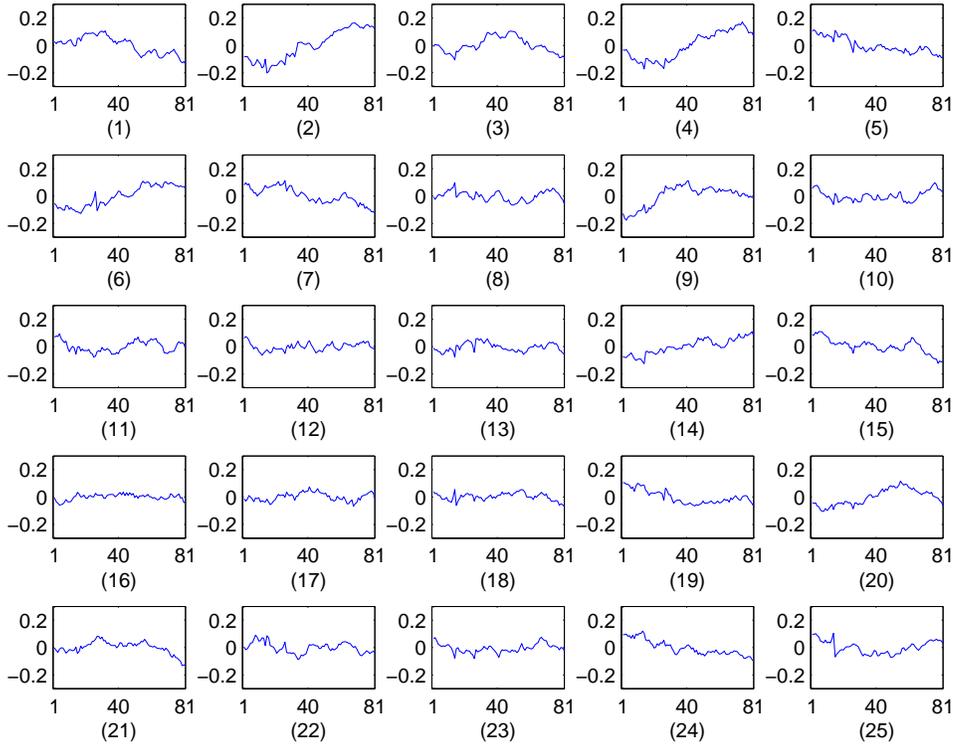}
\caption{A visualization of BSEs when the hidden layer has 25 nodes.}
\label{Fig:W8125}
\end{figure*}

Using SDSS training data (Section \ref{Sec:Data:SDSS}) and the scheme proposed in Section \ref{Sec:LearningBSE:Learning}, we learned 25 BSE structures (Fig. \ref{Fig:W8125}) by setting the number of hidden nodes $n = 25$ in Fig. \ref{Fig:autoencoder} and equation (\ref{Equ:BSEs}), where the optimality of n =25 are investigated in section \ref{Sec:Configuration}. It is shown that the BSE structure templates characterize the local structure of spectra. Using the learned BSE structures and the convolution-pooling scheme described in this section, we can extract features from every spectrum.

\begin{figure*}
  \centering
  \subfigure[]{
    \label{Fig:Con_pooling:sp} 
    \includegraphics[width =4in]{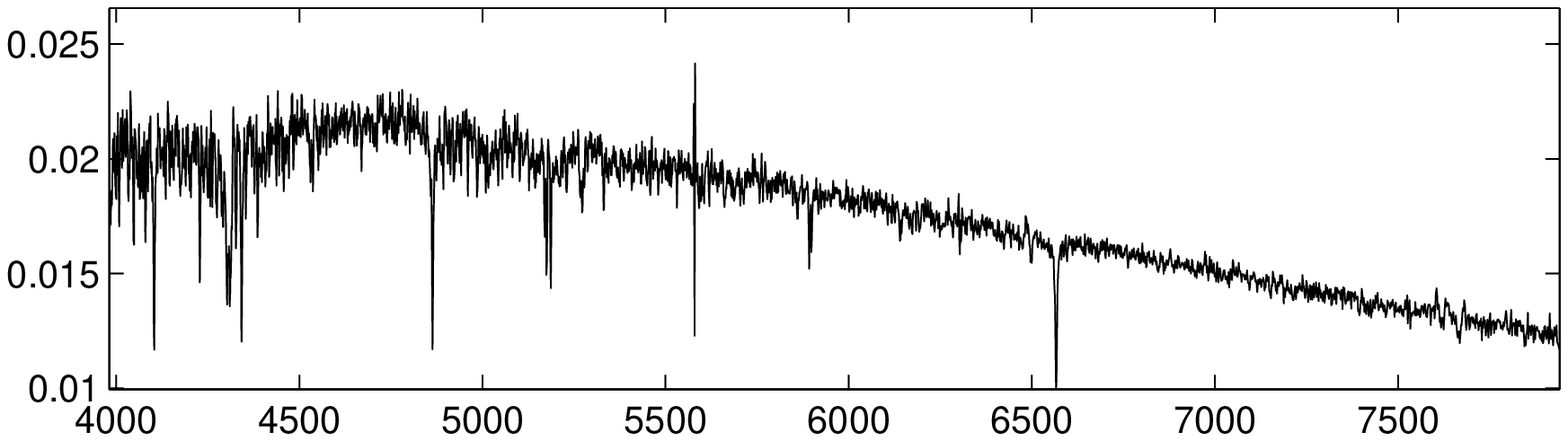}}
\hspace{-0.17in}
  \subfigure[]{
    \label{Fig:Con_pooling:Con} 
    \includegraphics[width =4in]{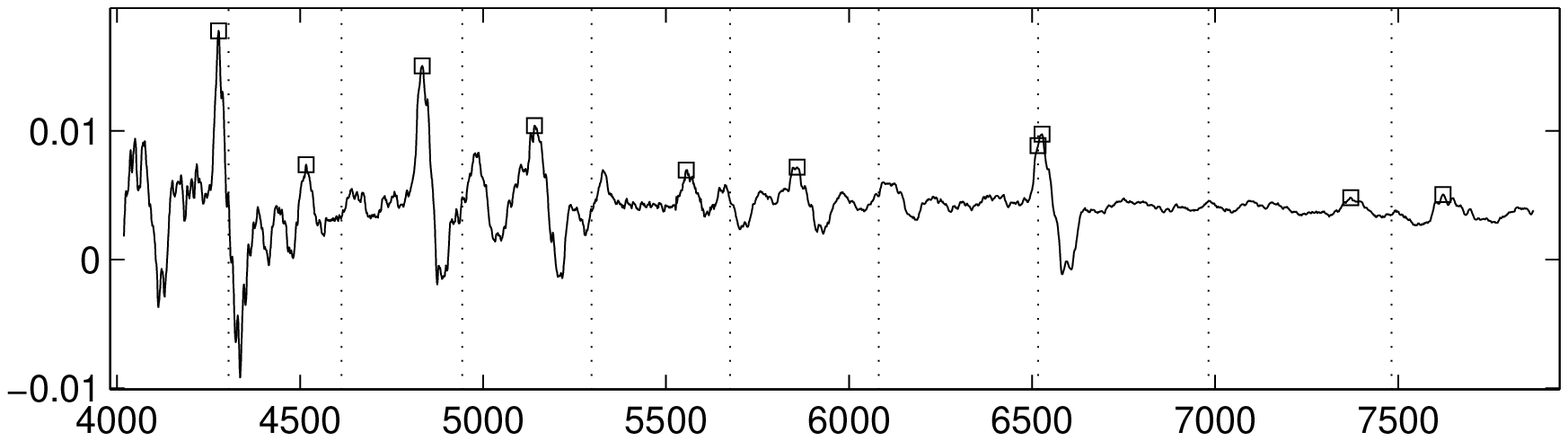}}
  \hspace{-0.17in}
  \subfigure[]{
    \label{Fig:Con_pooling:pooling} 
    \includegraphics[width =4in]{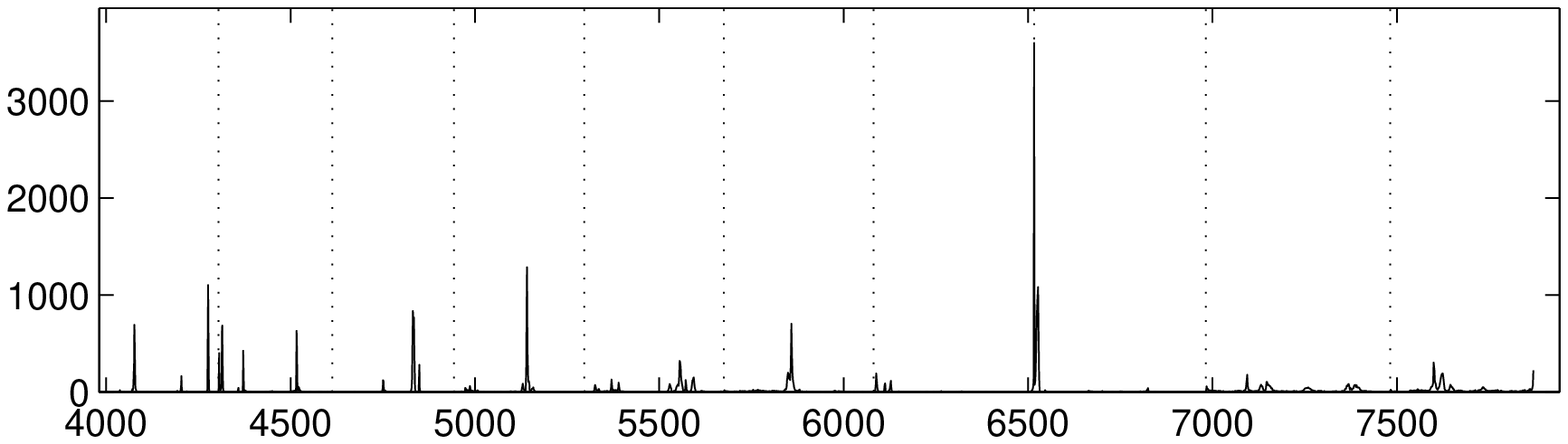}}
  \caption{Convolution responses of a spectrum and a histogram of the maximum convolution responses in a pooling process with pool number $N_p=10$.
  (a) A normalized SDSS spectrum.
  (b) Convolution responses of the spectrum shown in (a) and the third BSE shown in Fig. \ref{Fig:W8125}.
  In (b), nine vertical dashed lines are boundaries of each pool (the pooling is implemented in logarithmic wavelength space, so the sizes of multiple pools are unequal in a wavelength space) and ten quadrangles indicate the maximum convolution response in every pool.
  (c) The statistical histogram of the maximum convolution responses in the pooling process for the third BSE in Fig. \ref{Fig:W8125} on 5000 SDSS spectra (Section \ref{Sec:Data:SDSS}).
  }
  \label{Fig:maxcount1} 
\end{figure*}

For example, Fig. \ref{Fig:Con_pooling:sp} presents a spectrum from SDSS; there is a stitching error near 5580\AA. In Fig. \ref{Fig:Con_pooling:Con}, the curve shows the convolution response vector of the third BSE structure $W^{(1)}_{3\cdot}$ on the spectrum in Fig. \ref{Fig:Con_pooling:sp}; the points labeled with quadrangles are the pooling responses $\{v_q^{(3)}, q = 1, \cdots, N_p\}$ in equation (\ref{Equ:pooling:vec_sub}). The results in Fig. \ref{Fig:Con_pooling:Con} and \ref{Fig:Con_pooling:pooling} show that the responses of the spuriously strong signal of the stitching error are reduced.

\begin{figure*}
\centering
\includegraphics[width =4in]{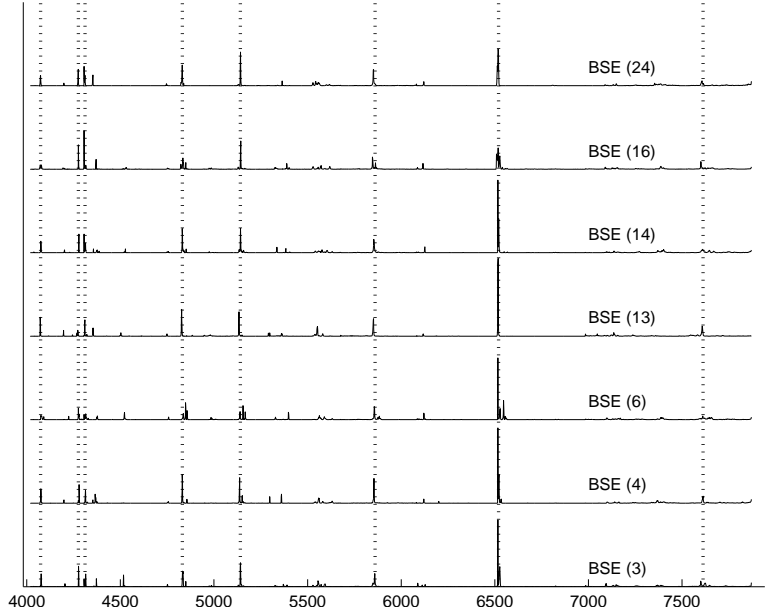}
\caption{ The cumulative histogram of the local maximum response position for the third, fourth, sixth, 13th, 14th, 16th, 24th BSE structures. The dashed lines indicate eight significant local maximum cumulative responses of pooling operation on 5000 SDSS spectra.}
\label{Fig:maxcount2} 
\end{figure*}

To investigate characteristics of the extracted features further using convolution and pooling, we calculate the statistical histogram of the maximum convolution responses on 5000 SDSS spectra (Section \ref{Sec:Data:SDSS}). Every maximum convolution response $v_q^{(j)}$ has a specific wavelength position (Fig. \ref{Fig:Con_pooling:Con}, equation \ref{Equ:pooling:vec_sub}). The statistical histogram of the maximum convolution responses is obtained by cumulating the number of pooling responses $v_q^{(j)}$ of 5000 SDSS spectra at each wavelength position. Fig. \ref{Fig:Con_pooling:pooling} shows the statistical histogram of the maximum convolution responses corresponding to the third BSE structure $W^{(1)}_{3\cdot}$ and more statistical histograms of the maximum convolution responses are presented in Fig. \ref{Fig:maxcount2}. The results in Fig. \ref{Fig:maxcount2} also show that the effects of stitching errors near 5580\AA~in Fig. \ref{Fig:Con_pooling:sp} are negligible in the pooling response. More experimental investigations on the stitching error are conducted in Section \ref{Sec:Experiments:stiching}.

The statistical histograms of the maximum convolution responses (Figs \ref{Fig:Con_pooling:pooling} and \ref{Fig:maxcount2}) on 5000 SDSS spectra also show that the pooling responses are statistically sparse: only at a few wavelength positions are there non-zero responses. The sparseness helps to trace back the physical interpretation of the extracted features.

That is to say, most of the cumulative responses are close to zero except a few wavelength patches with large cumulative responses. As can be seen in Fig. \ref{Fig:maxcount1}, these local maximum responses appear near the wavelength of evident fluctuation in a spectrum. In the pooling process, the responses of these wavelengths are frequently taken as spectral features. Some of the new features characterize such local structures in spectra and they are local and sparse.

Considering that a pooling response $v_q^{(j)}$ is obtained by a convolution calculation (equations \ref{Equ:pooling:vec_sub} and \ref{Equ:convolution:ele}), each pooling feature is actually calculated from some spectral fluxes in a wavelength range. Therefore, we present the wavelength ranges in the second column of Table \ref{Tab:Important Lines} for those features with large cumulative responses in Fig. \ref{Fig:maxcount2}.

Although the features are not extracted by identifying some spectral lines, the results in Fig. \ref{Fig:maxcount2} do show that the detected features are near some spectral lines potentially existing in a specific stellar spectrum (the third column of Table \ref{Tab:Important Lines}).

\begin{table}
\centering
\caption{ The wavelengths of the eight local-maximum cumulative responses and some potential lines near them. WPT: wavelength position of typical local-maximum cumulative response of a pooling operation on 5000 SDSS spectra, this position is represented by a three-dimensional vector $(a~~ b~~c)$, where $a$, $b$, $c$ are respectively
the starting wavelength, central wavelength and ending wavelength and $\log_{10}b = (\log_{10}a + \log_{10}c)/2$. PL: potential lines probably related to the description of the detected feature. Wavelength $b$ is the position of local maximum cumulative and fluxes from range $[a~~c]$ determine the response on b in a convolution process.
}
\begin{tabular}{ l l  c }
\hline
No.  &WPT({\AA})  &PL\\ \hline
1    &(4000 4074 4150)      &Ca\,{\sevensize I}, H\,{\sevensize $\delta$}, He\,{\sevensize I}\\ \hline
2    &(4199 4276 4356)   &Ca\,{\sevensize I}, H\,{\sevensize $\gamma$}\\ \hline
3    &(4233 4311 4391)   &Fe\,{\sevensize II}, Na\,{\sevensize II}, O\,{\sevensize I},Fe\,{\sevensize I},Ca\,{\sevensize III}\\ \hline
4    &(4744 4831 4922)   &Fe\,{\sevensize I}, O\,{\sevensize III}, Na\,{\sevensize I}, Na\,{\sevensize II}, O\,{\sevensize II}, Fe\,{\sevensize II}\\ \hline
5    &(5047 5141 5237)   &Fe\,{\sevensize II}, He\,{\sevensize I}, Na\,{\sevensize I},Ca\,{\sevensize III},, O\,{\sevensize III}, Fe\,{\sevensize I}\\ \hline
6    &(5757 5864 5973)   &Na\,{\sevensize I},Fe\,{\sevensize I},O\,{\sevensize II}\\ \hline
7    &(6401 6520 6641)   &H\,{\sevensize $\alpha$},CaH\\ \hline
8    &(7473 7612 7754)   &Fe\,{\sevensize I}, Fe\,{\sevensize II},O\,{\sevensize III},  O\,{\sevensize I}\\ \hline
\end{tabular}\label{Tab:Important Lines}
\end{table}

\section{Estimating Atmospheric Parameters}\label{Sec:Estimator}
As a typical non-linear learning method, ANN has been widely used in automated estimation of stellar atmosphere parameters \citep{Bailer-Jones2000, Willemsen2003, Ordonez2007, Ordonez2008, Zhao2008, Pan2009, Manteiga2010, Giridhar2011} and spectrum classification \citep{Gulati1994, Hippel1994, Vieira1995, Weaver1995, Schierscher2011}. For example, \citet{Manteiga2010} extracted spectral features using fast Fourier transforms (FFTs) and discrete wavelet transform (DWT) and estimated the parameters $T_\texttt{eff}$, log$~g$, [Fe/H] and [$\alpha$/Fe] by an ANN with one hidden layer from FFT coefficients and wavelet coefficients. \citet{Giridhar2011} studied how to estimate atmospheric parameters directly from spectra by a BP neural network. \citet{Pan2009} proposed to parameterize a stellar spectrum using an ANN from a few PCA features.

In this article, we use BP networks to learn the mapping from extracted DCP features (Section \ref{Sec:Feature Extract}) to stellar parameters $T_\texttt{eff}$, log$~g$ and [Fe/H]. Training of BP networks is an iterative process and in each iteration the estimated errors are calculated on two sets: the training set and the validation set. A BP network optimizes its parameters by minimizing the difference between the network's output and the expected output (e.g., stellar atmospheric parameters) according to the estimated errors in the training set and this process will stop when the estimated errors on the validation set have no improvement in successive iteration steps. This can avoid overfitting.

In a BP network, there are two preset parameters: the number of hidden layers and number of nodes in each hidden layer. For convenience, the former is denoted by $n^{BP}_{hl}$, the latter $\bm{n}^{BP}_{nhl}$. In this work, we investigated the cases $n^{BP}_{hl} = 1$ and $n^{BP}_{hl} = 2$ for computational feasibility. If
$n^{BP}_{hl} = 1$, $\bm{n}^{BP}_{nhl}$ is a positive integer. If $n^{BP}_{hl} = 2$, $\bm{n}^{BP}_{nhl}$ is a two-dimensional row vector consisting of the numbers of nodes in the first and second hidden layers.

Suppose that $S = \{(\bm{x},y)\}$ is a data set, where $\bm{x}$ represents the information of a spectrum and $y$ is an atmospheric parameter. In this work, the accuracy of a spectral parameterization $f(\cdot)$ on a data set $S$ is evaluated by the mean absolute error ($\texttt{MAE}$)
\begin{equation}\label{Equ:MAE}
    \texttt{MAE}(f(\cdot)) = \frac{1}{N} \sum_{(\bm{x},y) \in S}{\mid f(\bm{x}) - y \mid},
\end{equation}
where $N$ is the number of samples in $S$.

\section{Optimizing the configuration}\label{Sec:Configuration}

The proposed scheme consists of four steps (Fig. \ref{Fig:flowchart}, Section \ref{Sec:Framework}). In the second step, `Learning BSE', there are four parameters, $\lambda$, $\beta$, $\rho$ and $n$, to be preset, where $n$ denotes the number of nodes in the hidden layer of an autoencoder (Fig. \ref{Fig:autoencoder}). In the third step, `Extract features by convolution and pooling', there is a preset parameter, $N_p$, representing the number of pools (Section \ref{Sec:Feature Extract}). In the estimation method, BP network, there are two preset parameters $n^{BP}_{hl}$ and $\bm{n}^{BP}_{nhl}$ (Section \ref{Sec:Estimator}).

 To optimize the configuration of the proposed scheme, the spectrum-parameterization scheme was estimated from a training set from SDSS. The performance of the estimated spectrum-parameterization scheme was evaluated on a validation set from SDSS (Section \ref{Sec:Data}).

Therefore, the optimal configuration
\begin{equation}
(\hat{\lambda}, \hat{\beta}, \hat{\rho}, \hat{n}, \hat{N_p}, \hat{n}^{BP}_{hl}, \hat{\bm{n}}^{BP}_{nhl})_{ap}
\end{equation}
can be found by
\begin{equation}\label{Equ:config:obj}
\min\limits_{\lambda, \beta, \rho, n, N_p, n^{BP}_{hl}, \bm{n}^{BP}_{nhl}}{\texttt{MAE}(\lambda, \beta, \rho, n, N_p, n^{BP}_{hl}, \bm{n}^{BP}_{nhl}, ap)}
\end{equation}
where, $ap$ =$T_\texttt{eff}$, log$~g$ or [Fe/H] and $\texttt{MAE}$ is the prediction error on the SDSS validation set. The spectral parameterization is learned from a SDSS training set with a specific configuration of $\lambda$, $\beta$, $\rho$, $n$, $N_p$, $n^{BP}_{hl}$, $\bm{n}^{BP}_{nhl}$ and $ap$.

Because there is no any analytical expression for the objective function in equation (\ref{Equ:config:obj}), we can obtain the optional configuration $(\hat{\lambda}, \hat{\beta}, \hat{\rho}, \hat{n}, \hat{N_p}, \hat{n}^{BP}_{hl}, \hat{\bm{n}}^{BP}_{nhl})_{ap}$ in theory by repeating the four procedures of the proposed scheme (Fig. \ref{Fig:flowchart}, Section \ref{Sec:Framework}) with every possible configuration of $\lambda, \beta, \rho, n, N_p, n^{BP}_{hl}, \bm{n}^{BP}_{nhl}$, and choosing the one with minimal $\texttt{MAE}$ error as the optimal configuration.

However, this theoretical optimization scheme is infeasible as regards computational burden. Therefore, instead of obtaining an optimal configuration, we propose to find an excellent/acceptable configuration, a suboptimal solution.

To find a suboptimal solution, we restrict the search ranges for $\lambda, \beta, \rho, n$ and $N_p$ empirically, as follows:
\begin{equation}
\begin{split}
  RR_{\lambda} =& \{ 0.001,~0.0023,~0.0036,~0.0049,\\
                & ~0.0061,~0.0074,~0.0087,~0.01 \},
\end{split}
\end{equation}
\begin{equation}
\begin{split}
  RR_{\rho} =& \{ 0.005,~0.0114,~0.0179,~0.0243,\\
             &  0.0307,~0.0371,~0.0436,~0.05 \},
\end{split}
\end{equation}
\begin{equation}
  RR_{n} = \{ 15,~20, ~25, ~30, ~35, ~40 \},
\end{equation}
\begin{equation}
  RR_{N_p} = \{ 15, ~12, ~10, ~8, ~6, ~5, ~4\},
\end{equation}
and $\hat{\beta} =3$, where $RR_{\lambda}$ represents a restricted search range for $\lambda$; the other symbols $RR_{\rho}$, $RR_{n}$ and $RR_{N_p}$ are defined similarly. The suboptimal solutions of $\lambda, \rho, n$ and $N_p$ can be found by optimizing
\begin{equation}\label{Equ:config:obj:suboptimal}
\min\limits_{\lambda \in RR_{\lambda}, \rho \in RR_{\rho}, n \in RR_{n}, N_p \in RR_{N_p}}{\texttt{MAE}(\lambda, \hat{\beta}, \rho, n, N_p, \tilde{n}_{hl}, \tilde{\bm{n}}_{nhl}, ap)}
\end{equation}
based on the framework in Fig. \ref{Fig:flowchart}, where $n^{BP}_{hl}$ and $\bm{n}^{BP}_{nhl}$ are initialized with $\tilde{n}_{hl} =1$ and $\tilde{\bm{n}}_{nhl} = 6$. The $\tilde{n}_{hl}$ and $\tilde{\bm{n}}_{nhl}$ are determined empirically based on considerations of balance between computational burden and estimate accuracy. The configurations obtained are presented in Table \ref{Tab:suboptimal_configuration}.

Based on the configurations of $\lambda, \beta, n$ and $N_p$, the suboptimal solutions of  $n^{BP}_{hl}$ and $\bm{n}^{BP}_{nhl}$ are computed by
\begin{equation}\label{Equ:config:obj:suboptimal:2}
\min\limits_{n^{BP}_{hl} \in RR_{n^{BP}_{hl}}, \bm{n}^{BP}_{nhl} \in RR_{\bm{n}^{BP}_{nhl}}}{\texttt{MAE}(\hat{\lambda}, \hat{\beta}, \hat{n}, \hat{N_p}, n^{BP}_{hl}, \bm{n}^{BP}_{nhl}, ap)},
\end{equation}
where
$$
  RR_{n^{BP}_{hl}} = \{ 1, ~2 \}.
$$
If $n^{BP}_{hl} = 1$, then
$$
  RR_{\bm{n}^{BP}_{nhl}} = RR_{nn}.
$$
If $n^{BP}_{hl} = 2$, then
$$
  RR_{\bm{n}^{BP}_{nhl}} = \{ (a, ~b)| a \in RR_{nn}, b \in RR_{nn} \},
$$
where
\begin{equation}
  RR_{nn} = \{ 2, ~4, ~6, ~8, ~10, ~12, ~14, ~16, ~18, ~20, ~22, ~24 \}
\end{equation}
represents possible numbers of nodes in a hidden layer of a network investigated in this work.
The computed $(\hat{n}^{BP}_{hl}, ~\hat{\bm{n}}^{BP}_{nhl})$ are presented in Table \ref{Tab:suboptimal_configuration}.

\begin{table}
\centering
\caption{The suboptional configuration obtained for the proposed scheme (Fig. \ref{Fig:flowchart}, Section \ref{Sec:Framework}). $\hat{\lambda}$, $\hat{\beta}$, $\hat{\rho}$ and $\hat{n}$ are the optimized values of four parameters in the second step, `Learning BSE' (Fig. \ref{Fig:flowchart}), where $\hat{n}$ denotes the number of nodes in the hidden layer (Fig. \ref{Fig:autoencoder}).
$\hat{N}_p$ represents the number of pools (Section \ref{Sec:Feature Extract}) in the third step, `Extract features by convolution and pooling' (Fig. \ref{Fig:flowchart}).
$\hat{n}^{BP}_{hl}$ and $\bm{\hat{n}}^{BP}_{nhl}$ are the optimized values of two preset parameters in the estimation method, BP network (Section \ref{Sec:Estimator}).}
\begin{tabular}{ c l  l  l  l  l  l l}
  \hline
Parameters            &$\hat{\lambda}$   &  $\hat{\beta}$  & $\hat{\rho}$  & $\hat{n}$                 & $\hat{N}_p$ &  $\hat{n}^{BP}_{hl}$ &$\hat{\bm{n}}^{BP}_{nhl}$ \\ \hline
log$~T_\texttt{eff}$  &0.0074            &3                &0.0050   &   25                     &10             & 2           & (14,10 )\\ \hline
log$~g$               &0.0087            &3                & 0.114   &   25                     &12             & 2           &(16,4)\\ \hline
[Fe/H]                &0.0087            &3                & 0.114   &   25                     &15             & 2           &(22,6)\\ \hline
\end{tabular}\label{Tab:suboptimal_configuration}
\end{table}

Note, however, that the restricted search ranges are selected empirically based on performance on validation set, the solutions obtained are not usually global minimum ones, but some acceptable values with a feasible computational burden. If we expand the restricted ranges, it is possible that a better configuration can be obtained at greater computational cost.

\section{Experiments and Discussion}\label{Sec:Experiments}

\subsection{Performance on SDSS spectra}\label{Sec:Experiments:SDSS}
From the training set and the validation set consisting of SDSS spectra (Section \ref{Sec:Data}), we obtain a spectral parameterization using the proposed scheme (Section \ref{Sec:Framework}, Fig. \ref{Fig:flowchart}) and the proposed optimization scheme (Section \ref{Sec:Configuration}, Table \ref{Tab:suboptimal_configuration}).

On a SDSS test set, the $\texttt{MAE}$ errors of this spectral parameterization are 0.0060 dex for log$~T_\texttt{eff}$, 0.1978 dex for log$~g$ and 0.1770 dex for [Fe/H]
(the row with index 1 in Table \ref{Tab:performance_eva}). Similarly, on real spectra from SDSS, \citet{Journal:Fiorentin:2007} investigated the stellar parameter estimation problem using PCA and ANN and obtained accuracies of 0.0126 dex for log$~T_\texttt{eff}$, 0.3644 dex for log$~g$ dex and 0.1949 dex for [Fe/H] on a SDSS test set based on 50 PCA features. \cite{Liu2014} studied the spectrum-parameterization problem using least absolute shrinkage and selection operator (LASSO) algorithm and Support Vector Regression (SVR) method and the optimal $\texttt{MAE}$ errors are 0.0094 dex for log$~T_\texttt{eff}$, 0.2672 dex for log$~g$ and 0.2728 dex for [Fe/H].

\begin{table*}
\centering
\caption{Performance of the proposed scheme on test data sets}
\begin{tabular}{ c c c c c}
\hline \hline
Index &Data source            &  log~$T_\texttt{eff}$ (dex)  &   log$~g$ (dex)    &   [Fe/H](dex)  \\ \hline
\multicolumn{5}{c}{(a) Performance of the proposed scheme. More details are presented in Section \ref{Sec:Experiments:SDSS} and \ref{Sec:Experiments:synthetc}. } \\
\hline
1     &SDSS spectra           &  0.0060               &       0.1978         &   0.1770\\ 
2     &synthetic spectra      &  0.0004               &       0.0145         &   0.0070\\ \hline
\multicolumn{5}{c}{(b) Rationality to delete some data components in application (Section \ref{Sec:Experiments:balance}). } \\
\hline
3     &SDSS spectra    &  0.0080               &       0.2994         &   0.2163\\ \hline
\multicolumn{5}{c}{(c) Robustness to stitching error (Section \ref{Sec:Experiments:stiching}). } \\ \hline
4     &SDSS spectra         &0.0063   &0.2371  &0.1827\\
 \hline \hline
\end{tabular}\label{Tab:performance_eva}
\end{table*}

\subsection{Performance on synthetic spectra}\label{Sec:Experiments:synthetc}
To investigate the effectiveness of the proposed scheme further, we also evaluated it on synthetic spectra. The synthetic spectra used in this work and those used in \citet{Journal:Fiorentin:2007} are all calculated from Kurucz's model. The synthetic data set is described in Section \ref{Sec:Data:synthetic}. This experiment shares the same parameters as the experiment on SDSS data (Section \ref{Sec:Experiments:SDSS}) and the BP estimation is learned from the synthetic training set (Section \ref{Sec:Data}).

On the synthetic test set, the $\texttt{MAE}$ accuracies of the spectral parameterization learned from the synthetic training set are 0.0004 dex for log$~T_\texttt{eff}$, 0.0145 dex for log$~g$, and 0.0070 dex for [Fe/H] (the row with index 2 in Table \ref{Tab:performance_eva}). In \citet{Journal:Fiorentin:2007}, the best consistencies on synthetic spectra are obtained based on 100 principal components and the $\texttt{MAE}$s are 0.0030 dex for log$~T_\texttt{eff}$, 0.0251 for log$~g$ and 0.0269 for [Fe/H](Table 1 in \citet{Journal:Fiorentin:2007}). Using the LASSO algorithm and SVR method, \citet{Journal:Li:2014} reached $\texttt{MAE}$ errors 0.0008 dex for log$~T_\texttt{eff}$, 0.0179 dex for log$~g$ and 0.0131 dex for [Fe/H]. On the synthetic spectrum, the mean absolute errors in \cite{Manteiga2010} are 0.07 dex  for log$~g$ and 0.06 dex for [Fe/H]. 

\subsection{Effective data components, unwanted influences and their balances}\label{Sec:Experiments:balance}
The proposed scheme extracts features by throwing away some information. In this procedure, it is very probable that some useful (at least in theory) components are discarded. Is this positive or negative?

Actually, besides the useful data components in spectra, there is also redundancy, and/or noise and pre-processing imperfections (e.g. sky lines and/cosmic ray removal residuals, residual calibration defects). Redundancy means that some duplication of some components in a system exists. Multiple components are probably usually different from each other regarding the amount of duplications. Therefore, redundancy can disturb the weights of different components, which usually results in an erroneous evaluation and reduces the quality of learning. Noise and pre-processing imperfections can mask off the effects of some important spectral information, e.g. weak lines. Therefore, in theory, it is possible that we can improve the estimates of atmospheric parameters by throwing away some data components.

We conducted an experiment to investigate this theoretical possibility. In this experiment, we estimate atmospheric parameters by deleting procedures 2 and 3 (Fig. \ref{Fig:flowchart}) from the experiments in Section \ref{Sec:Experiments:SDSS}. That is to say, we estimated the atmospheric parameters using a BP network directly from a spectrum without throwing away any information in the pooling procedure and the BP estimation network shared the same configurations as in the experiment of Section \ref{Sec:Experiments:SDSS}. The results are presented in
Table \ref{Tab:performance_eva} (b) (the row with index 3).
 The results of the experiments with index 1 and 3 in Table \ref{Tab:performance_eva}
 do not show any evident evidence of losing any spectrum parameterization performance when throwing away some data components using the proposed scheme.

\subsection{Effects of band stitching errors on parameter estimation}\label{Sec:Experiments:stiching}
There is a significant oscillation in some SDSS spectra near 5580 {\AA} (e.g. Fig. \ref{Fig:Con_pooling:sp}), which is caused by errors in stitching the red and blue bands. Figs \ref{Fig:Con_pooling:Con}, \ref{Fig:Con_pooling:pooling} and \ref{Fig:maxcount2} show that the responses of stitching errors is evidently reduced in the procedures of convolution and pooling.

We also performed one experiment to analyse the effects of these `noises' on performances of the proposed scheme. In this experiment, the BP estimator is trained with features excluding the convolution responses related to stitching errors. For example, in SDSS data, the stitching error appears in the fifth pool in our experiments, so we removed the pooling response of the fifth pool for every BSE structure. The new $\texttt{MAE}$s of the three parameters in the SDSS test set are presented in Table \ref{Tab:performance_eva} , part (c).

Actually, in the wavelength range containing stitching error, there are both useful components and disturbances from noise and stitching errors. However, experiments show that the useful components outperform the disturbances in the proposed scheme (the experiments with indexes 1 and 4 in Table \ref{Tab:performance_eva}).

\section{Conclusion}\label{Sec:Conclusion}
In this work, we propose a novel scheme for spectral feature extraction and stellar atmospheric parameter estimation. In the commonly used methods like PCA and Fourier transform, every feature is obtained by incorporating nearly all of the fluxes of a spectrum. Differently from these, the characteristics of our proposed scheme are localization and sparseness of the extracted features. `Localization' means that each extracted feature is calculated from some spectral fluxes within a local wavelength range (Fig. \ref{Fig:Con_pooling:Con}). `Sparseness' says that the atmospheric parameters can be estimated using a small number of features (Fig. \ref{Fig:Con_pooling:pooling}). The `localization' and `sparseness' signify that many data components are thrown away, especially the redundancies, weak informative components (Section \ref{Sec:Experiments:balance}). However, these weak informative components are easily corrupted by noise and pre-processing imperfections (Section \ref{Sec:Experiments:balance}). It is shown that the proposed scheme has excellent performance in estimating atmospheric parameters (Section \ref{Sec:Experiments:SDSS}, Section \ref{Sec:Experiments:synthetc}, Table \ref{Tab:performance_eva} (a)
).

\section*{Acknowledgments}

The authors thank the reviewer and editor for their instructive comments and extend their thanks to
Professor Ali Luo for his support and invaluable discussions.
This work is supported by the National Natural Science Foundation of China (grant No: 61273248, 61075033, 61202315), the Natural Science Foundation of Guangdong Province (2014A030313425,S2011010003348), the Open Project Program of the National Laboratory of Pattern Recognition(NLPR) (201001060) and the high-performance computing platform of South China Normal University.

\appendix
\bsp
\label{lastpage}
\end{document}